\documentclass[12pt]{article}
\usepackage{amsmath,amssymb,color}
\usepackage{cite,epsf}
\usepackage{pstricks}
\usepackage{color}

\usepackage{graphicx,array} 
\newcommand{\be}{\begin{equation}}
\newcommand{\ee}{\end{equation}}
\newcommand{\beq}{\begin{equation}}
\newcommand{\eeq}{\end{equation}}
\newcommand{\bea}{\begin{eqnarray}}
\newcommand{\eea}{\end{eqnarray}}

\newcommand{\ba}{\begin{eqnarray}}
\newcommand{\ea}{\end{eqnarray}}

\def\sin{\mbox{sin}}
\def\cos{\mbox{cos}}
\def\cot{\mbox{cot}}
\def\log{\mbox{log}}

\begin{document}

\begin{titlepage}
\vspace{10pt}
\hfill
{\large\bf HU-EP-16/05}
\vspace{20mm}
\begin{center}

{\Large\bf  Holographic entanglement entropy \\[2mm]
for hollow cones and banana shaped regions  \\[2mm]

}

\vspace{45pt}

{\large Harald Dorn 
{\footnote{dorn@physik.hu-berlin.de
 }}}
\\[15mm]
{\it\ Institut f\"ur Physik und IRIS Adlershof, 
Humboldt-Universit\"at zu Berlin,}\\
{\it Zum Gro{\ss}en Windkanal 6, D-12489 Berlin, Germany}\\[4mm]

\vspace{20pt}

\end{center}
\vspace{10pt}
\vspace{40pt}

\centerline{{\bf{Abstract}}}
\vspace*{5mm}
\noindent
We consider banana shaped regions as examples of compact regions, whose boundary has  two conical singularities. Their
regularised holographic entropy is calculated with all divergent as well as finite terms. 
The coefficient of the squared logarithmic divergence, also in such a case with internally curved boundary, agrees with that calculated in the literature
for infinite circular cones with their internally flat boundary. For the otherwise conformally invariant coefficient of the ordinary logarithmic divergence an anomaly under exceptional conformal transformations is observed. 

The construction of minimal submanifolds, needed for the entanglement entropy of cones, requires fine-tuning of Cauchy data. Perturbations of such fine-tuning leads
to solutions relevant for hollow cones. The divergent parts
for the entanglement entropy of hollow cones are calculated. Increasing the difference
between the opening angles of their outer and inner boundary, one finds
a transition between connected solutions for small differences to disconnected solutions
for larger ones. 

\vspace*{4mm}
\noindent

\vspace*{5mm}
\noindent
   
\end{titlepage}
\newpage


\section{Introduction}
For a quantum field theory in $(d+1)$ dimensions the entanglement entropy of a $d$-dimensional spatial region
${\cal A}$ is defined by \footnote{For reviews see e.g. \cite{Ryu:2006ef},\cite{Solodukhin:2011gn} and refs. therein.}
\beq
S({\cal A})~=~-\mbox{tr}(\rho_{\cal A}~\log\rho_{\cal A})~,
\eeq
where $\rho _{\cal A}$ is the density operator obtained by integrating out the degrees of freedom outside ${\cal A}$.
In conformal field theories with a holographic dual, defined in $AdS_{d+2}$ or some of its modifications, $S({\cal A})$ at strong coupling can be related
to the volume of the minimal spatial $d$-dimensional submanifold $\gamma_{\cal A}\subset AdS_{d+2}$, which  approaches the boundary $\partial {\cal A}$
of ${\cal A}$ on the boundary of $AdS$, by \cite{Ryu:2006bv,Ryu:2006ef}
\beq
S({\cal A})~=~\frac{V(\gamma_{\cal A})}{4~G_N^{(d+2)}}~.\label{holo}
\eeq 
In this formula $V(\gamma_A)$ is the volume of $\gamma_{\cal A}$ and $G_N^{(d+2)}$ denotes the $(d+2)$-dimensional Newton constant.

Due to the near boundary behaviour of the $AdS$ metric, these volumes are divergent. The standard procedure
for a regularisation refers to the use of  Poincar{\' e}  coordinates
\beq
ds^2~=~\frac{1}{r^2}~\big ( dr ^2 ~+~dx^{\mu}dx_{\mu}\big )\label{poincare}
\eeq
and cutting off that part of the submanifold whose $r$-coordinate is smaller than $\epsilon$.  
For smooth $\partial {\cal A}$, which by construction is the boundary of $\gamma_{\cal A}$, the small $\epsilon$-expansion of $V_{\epsilon}(\gamma_{\cal A})$ has the
following structure \cite{Graham:1999pm} 
\bea
V_{\epsilon}&=&\frac{c_1}{\epsilon^{d-1}}~+~\frac{c_3}{\epsilon^{d-3}}~+~\dots ~+~\frac{c_{d-2}}{\epsilon^2}~+~a~\log\epsilon~+~c_d~+~o(1)~,~~~\mbox{for odd}~d~, 
\label{odd-GW}\\[2mm]
V_{\epsilon}&=&\frac{c_1}{\epsilon^{d-1}}~+~\frac{c_3}{\epsilon^{d-3}}~+~\dots ~+~\frac{c_{d-1}}{\epsilon}~+~c_d~+~o(1)~,~~~\mbox{for even}~d~.~~~~~~~~~~~~
\eea
Although conformal transformations on the boundary of $AdS$ act as isometries in the bulk, conformal invariance
is broken by the use of the cut-off. Nevertheless  for odd $d$ the special coefficient $a$ 
and for even $d$ the special coefficient $c_d$ are invariant with respect to conformal transformations of
the boundary of $\gamma _{\cal A}$ \cite{Graham:1999pm}. 

The coefficient of the leading divergence is proportional to the volume of 
$\partial {\cal A}$ \cite{Ryu:2006ef} and
the coefficient of the $\log\epsilon$ term can be expressed as an integral over a  conformally invariant
quantity constructed  out of the second fundamental form of $\partial {\cal A}$ \cite{Solodukhin:2008dh}. 

For singular boundaries additional divergences show up. In particular, for isolated conical singularities
of $\partial {\cal A}$ (cusps in the case $d=2$) these new contributions are logarithmic for even $d$ and double logarithmic
for odd $d$ \cite{Myers:2012vs}. Their behaviour in the smooth limit has been related to a certain  central charge of the CFT \cite{Bueno:2015lza}. The minimal surfaces needed for the entanglement
entropy in $d=2$ are also relevant for the holographic treatment of the strong coupling behaviour of Wilson loops  \cite{Rey:1998ik,Maldacena:1998im}. There the 
coefficient of the logarithmic divergence, called cusp anomalous dimension, has been calculated in \cite{Drukker:1999zq}. The corresponding discussion applied to the entanglement entropy can be found in \cite{Hirata:2006jx,Hirata:2008ms}.\\

As in the two-dimensional case, also in  higher dimensions the extraction of the coefficient of the additional divergence, generated by a conical singularity, has been
performed by choosing $\partial {\cal A}$ as the boundary of an infinite circular cone \cite{Klebanov:2012yf,Myers:2012vs}. It is then expected, that the coefficient remains 
unchanged if one allows arbitrary curvature in the neighbourhood
of the singular point, while keeping the opening angle fixed. We have checked
this expectation for the two-dimensional case \cite{Dorn:2015bfa}, both by
the calculation of the area of the minimal surface related to a curved compact $\partial {\cal A}$ obtained by two intersecting circles, as well as by a general proof. 

The present paper is devoted to the analogous problem for $d=3$. We will calculate the
regularised volume up to terms vanishing for $\epsilon\rightarrow 0$ for the three-dimensional minimal submanifold $\gamma _{\cal A}$,
which reaches the boundary of Euclidean $AdS_4$ \footnote{The time coordinate in $AdS_5$ is fixed.} and meets there the boundary of a banana shaped region ${\cal A}$.
\footnote{To our knowledge it will be the first explicit calculation
for a {\it compact} region ${\cal A}$ with singularities on its boundary $\partial {\cal A}$.}
The construction is technically a bit more involved as in $d=2$, since a helpful
conservation law is no longer available, and one has to handle a second order 
 nonlinear differential equation instead of an integrable first order one. As a
by-product of our analysis we will find the coefficients of divergences
for a new type of singularity of $\partial {\cal A}$, the hollow cone.

The paper is organised as follows. To set up some notation we review and comment 
in section 2 the calculation for the infinite circular cone as performed in \cite{Klebanov:2012yf,Myers:2012vs}. In section 3 we study the issue of stability under
perturbations of the Cauchy initial data for the differential equation under study.
This will bring us solutions for hollow cones, whose inner and outer conical surfaces have a common tip.

Section 4 collects some elementary geometrical properties of  certain banana shaped
regions, and in section 5 we apply a suitable conformal transformation of cones to get
the regularised volume $V_{\epsilon}(\gamma_{\cal A})$ for banana shaped regions ${\cal A}$. We conclude in section 6 and have put some technical details related to section 5 into appendix A.
The second appendix B presents facts on an anomaly under exceptional conformal
transformations needed in the conclusion section.
\section{Entanglement entropy for a cone}
For the holographic entanglement entropy of a cone in ${\mathbb R}^3$ the divergent parts of the relevant
volume have been calculated in \cite{Myers:2012vs,Klebanov:2012yf}. Here we repeat
some steps of the calculation, both to set up some notation and to   pick up also the finite part.

One has to find the 3-dimensional minimal submanifold
 in the bulk with Euclidean $AdS_4$-metric in Poincar{\' e}  coordinates, see \eqref{poincare},
which approaches the boundary of the cone for $r\rightarrow 0$. 
Using the symmetry of the problem we can make the ansatz
\bea
x_1&=&\rho ~\sin\vartheta~\cos\varphi~,~~~x_2~=~\rho~\sin\vartheta~\sin\varphi~,~~~x_3~=~\rho ~\cos\vartheta~,\nonumber\\
r&=&\rho~h(\vartheta )~,\label{mf}
\eea
with $0\leq\rho<\infty~,~~0\leq\varphi<2\pi ~,~~0\leq \vartheta\leq \Omega $ and $2\Omega$ denoting the opening
angle of the cone ($0<\Omega<\pi/2$). The boundary condition for $h(\vartheta)$ is
\beq
h(\Omega)~=~0~.\label{bc}
\eeq
The volume of the manifold \eqref{mf} is ($\dot h=dh(\vartheta)/d\vartheta $)
\beq
V~=~2\pi\int _0^{\infty}\frac{d\rho}{\rho}\int_0^{\Omega} d\vartheta ~\frac{\sin\vartheta}{h^3(\vartheta)}\sqrt{1+h^2+\dot h^2}~.\label{lagrangian}
\eeq 
It needs both an IR regularisation at large $\rho$ as well as an UV regularisation near the boundary of $AdS$, i.e $\vartheta =\Omega$ or $\rho =0$.

In the analogous case in $AdS_3$ there is no explicit $\vartheta$-dependence in the integrand, and the related
conservation law yields a first order differential equation which can be solved by integration \cite{Drukker:1999zq}. 

However, here $\vartheta$ appears explicitly, and we are forced to handle the second order differential equation
which enforces the stationarity condition for \eqref{lagrangian}
\bea
\big (~\ddot h (h+h^3)+\dot h^2(3+h^2)+3+5h^2+2h^4~\big )&\sin\vartheta & \nonumber\\[2mm]
+~h\dot h(1+h^2+\dot h^2)&\cos\vartheta &=~0~.\label{eom}
\eea
This equation is singular at $\vartheta =0$. If one looks for regular solutions
at this point, they have to obey either $\dot h=0$ or $h=0$, implying that 
there the initial value problem cannot be solved for generic Cauchy data.
Furthermore, there appear movable singularities, whose positions depend
on the initial conditions. In particular, any zero of $h$ in $(0,\pi/2)$
can occur only in combination with diverging derivatives.
 
\begin{figure}[h!]
 \centering
 \includegraphics[width=9cm]{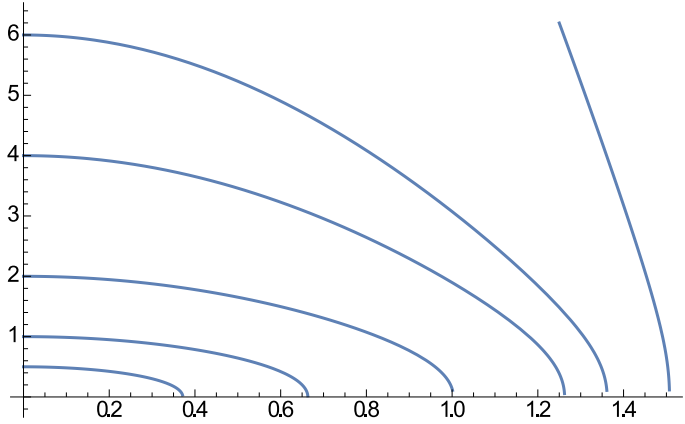} 
\caption{\it Some examples for the function $h(\vartheta)$, obtained as 
numerical solutions
$~~~~~~~~~~~~~~$ of the differential equation \eqref{eom} with boundary condition \eqref{bc} and finite $h(0)$.} 
\label{num-sol}
\end{figure}

The solutions regular at $\vartheta =0$ have a power series expansion in $\vartheta ^2$
\beq
h(\vartheta)~=~h_0~-~\frac{3+2h_0^2}{4h_0}~\vartheta ^2 ~+~\frac{16h_0^6-32h_0^4-174h_0^2-135}{384h_0^3(1+h_0^2)}~\vartheta^4~+~{\cal O}(\vartheta ^6)~.
\eeq
Some examples of numerical solutions are shown in fig.\ref{num-sol}. Obviously, $h(\vartheta)$ has its maximum
value $h_0$ at $\vartheta =0$ and falls monotonically to zero at $\vartheta =\Omega $.
 The relation
between $h_0$ and $\Omega $ is one to one, the function $h_0(\Omega)$ is monotonically increasing.

For our further analysis it is sufficient to solve \eqref{eom} with boundary condition \eqref{bc}
near $\vartheta =\Omega$. We get
\bea
h(\Omega -\delta )&=&2 (\tan\Omega)^{1/2}~\delta ^{1/2} ~+~\frac{(3-\cos(2\Omega))(\cot\Omega )^{1/2}}{8~\cos ^2\Omega}~\delta ^{3/2}\log \delta \label{h-expanded}\\
&&+~h_*~\delta ^{3/2}~+~{\cal O}(\delta ^{5/2}\log^2\delta )~.\nonumber
\eea
If one looks for arbitrary solutions of \eqref{eom} approaching $h=0$, the constant $h_*$ remains free. \footnote{We have checked explicitly, that in the asymptotic series
for the solution of \eqref{eom} the coefficients of the next order terms $\delta ^{5/2}\log^n\delta~,~n=0,1,2$ can be expressed in terms of $\Omega$ and $h_*$.}
For those solutions, depicted in fig.\ref{num-sol}, which start with $h(0)=h_0$ and smoothly fall down to $h=0$, the constant $h_*$ is a function of $h_0$.
Note also, that due to \eqref{h-expanded} real solutions
can approach $h=0$ at $\vartheta =\Omega\in (0,\pi/2) $ only from the side of
lower $\vartheta$-values.
 
With \eqref{h-expanded} we can confirm a formula taken out of \cite{Myers:2012vs}
\beq
\sin\vartheta ~=~\sin\Omega ~-~\frac{\cos\Omega~\cot\Omega }{4}~h^2~+~\frac{\big (3-\cos(2\Omega)\big )~\cot ^2\Omega}{64~\sin\Omega}~h^4~\log h~+~{\cal O}(h^4)\label{sintheta}~.
\eeq
Since $h(\vartheta)$ is monotonically decreasing, we can use the inversion $\vartheta (h)$ to change the 
integration variable in \eqref{lagrangian} from $\vartheta$ to $h$. Defining
\beq
F(h)~=~\frac{\sin\vartheta(h)}{h^3\dot h(\vartheta (h))}\sqrt{1+h^2+\dot h^2}\label{Fh}
\eeq
one gets \cite{Myers:2012vs}
\beq
F(h)~=~-\frac{\sin\Omega}{h^3}~+~\frac{\cos\Omega ~\cot\Omega}{8h}~+~{\cal O}(h)~.
\label{Fas}
\eeq

Following the standard UV-regularisation by cutting off that part of the manifold whose distance 
in the Poincar{\' e}  coordinate $r$ is smaller than $\epsilon$ and using the IR-cut-off $\rho <l$ we get
for the regularised volume of \eqref{mf}
\beq
V_{\epsilon,l}~=~~2\pi\int_{h_0}^{\epsilon /l} dh ~F(h) \int _{\epsilon/h}^{l}\frac{d\rho}{\rho}~.
\eeq
To handle the divergence of $F(h)$ at $h\rightarrow 0$ we define its finite piece $\tilde F$ by
\beq
F(h)~=~\tilde F(h)~-\frac{\sin\Omega}{h^3}~+~\frac{\cos\Omega ~\cot\Omega}{8h}~.\label{Ftilde}
\eeq
Then from \eqref{Fas} we know $\tilde F~=~{\cal O}(h)$.

Performing the $\rho$-integration one gets
\beq
V_{\epsilon,l}~=~2\pi\int^{h_0}_{\epsilon/l}dh\Big (\tilde F(h)-\frac{\sin\Omega}{h^3}+\frac{\cos\Omega~\cot\Omega}{8h}\Big )~\log\frac{\epsilon}{lh}~.
\eeq
In the limit $\epsilon\rightarrow 0$ in the part of the integral with $\tilde F$ one can replace the lower boundary by zero. The remaining parts of the $h$-integral can be performed explicitly. The final result is
\bea
V_{\epsilon,l}&=&\frac{\pi ~\sin\Omega}{2}~\frac{l^2}{\epsilon^2}~-~\frac{\pi~\cos\Omega~\cot\Omega}{8}~\log^2~\frac{\epsilon}{l}~+~o(1)\\[2mm]
&&+~\Big (2\pi\int _0^{h_0}\tilde F(h)dh~+~\frac{\pi~\sin\Omega}{h_0^2}~+~\frac{\pi~\cos\Omega~\cot\Omega}{4}~\log h_0~\Big )~\log\frac{\epsilon}{l}\nonumber\\[2mm]
&&-~2\pi\int_0^{h_0}dh ~\tilde F~\log h~-~\frac{\pi (1+2~\log h_0)~\sin\Omega}{2h_0^2}~-~\frac{\pi~\cos\Omega~\cot\Omega}{8}~\log ^2h_0~.\nonumber
\eea 
The coefficient of the most singular term $1/\epsilon ^2$ is equal to half of the area of the boundary of the cone cut at $\rho =l$.

Some further observations on the stability issue of these solutions, which go beyond the discussion in \cite{Myers:2012vs}, are presented in the next section.

\section{Stability analysis for the cone solutions and study of hollow cone solutions}
\begin{figure}[h!]
 \centering
 \includegraphics[width=10cm]{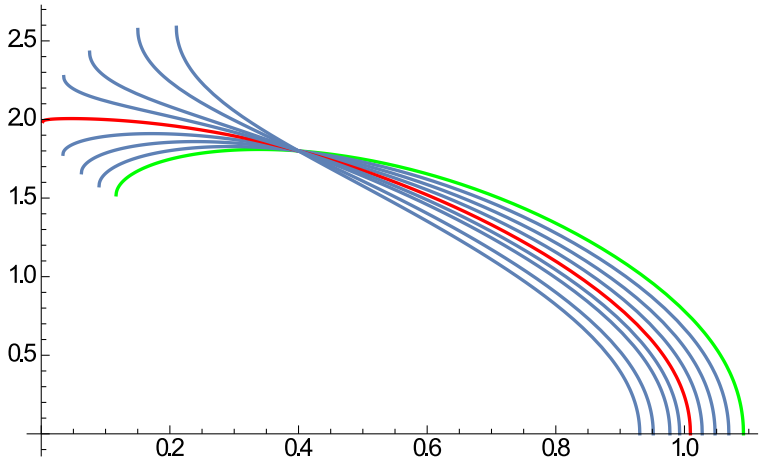} 
\caption{\it Numerical solutions $h(\vartheta)$ of the differential equation \eqref{eom}, obtained
by solving the Cauchy initial problem at $\vartheta =0.4$ with $h=1.8$ and different values for $\dot h$. The red curve in the middle corresponds to an approximation for a solution reaching its maximum $h_0$ at $\vartheta =0$. The green marking is made for reference to fig.\ref{num-sol3}.} 
\label{num-sol2}
\end{figure}

Depending on the initial data, the numerical evaluation of \eqref{eom} stops at some
points $(\vartheta,h)$ with $\vartheta,~h\geq 0$ where  $\dot h(\vartheta)\rightarrow\pm\infty$, see fig.\ref{num-sol2}.
We expect that this is an artefact of the chosen parameterisation of the minimal submanifold.
It continues smoothly beyond such a point, just in a manner that
the inverse function $\vartheta (h)$ has a local minimum. To prove this statement, we transform 
the differential equation \eqref{eom} in one for the inverse function and get ($'$ denoting $d/dh$)
\beq
\vartheta ''~=~\frac{1+\vartheta '^2(1+h^2)}{1+h^2}~\cot\vartheta(h) ~+~\frac{\vartheta '}{h}~\frac{3+h^2+\vartheta'^2(3+5h^2+2h^4)}{1+h^2}~.
\eeq
Obviously, points with $h\neq 0$ and $0<\vartheta<\pi$ are regular, allowing solutions
of the full Cauchy initial value problem. With $\vartheta'=0$ (corresponding to $\dot h=\pm\infty$) 
one arrives at
\beq
\vartheta''~=~\frac{\cot\vartheta}{1+h^2}~,
\eeq
showing that for the  function $\vartheta(h)$ local extrema can appear in the strip  $0<\vartheta<\pi/2$ ($h=0$ still excluded) as minima only and in the strip  $\pi /2<\vartheta<\pi$ as maxima only. 
In a similar way one gets from the original differential equation \eqref{eom}, that
in $0<\vartheta<\pi$ at local extrema $\ddot h/h<0$, excluding therefore local minima of $h(\vartheta)$ for $h>0$ and local maxima for $h<0$. This gives us already a lot of information on the smooth continuations
of the curves in fig.\ref{num-sol2}. Concerning the vertical direction only turns 
into the direction of the $\vartheta$-axis are allowed. In horizontal direction only
turns into the direction of the line $\vartheta=\pi/2$ can show up.

\begin{figure}[h!]
 \centering
 \includegraphics[width=14cm]{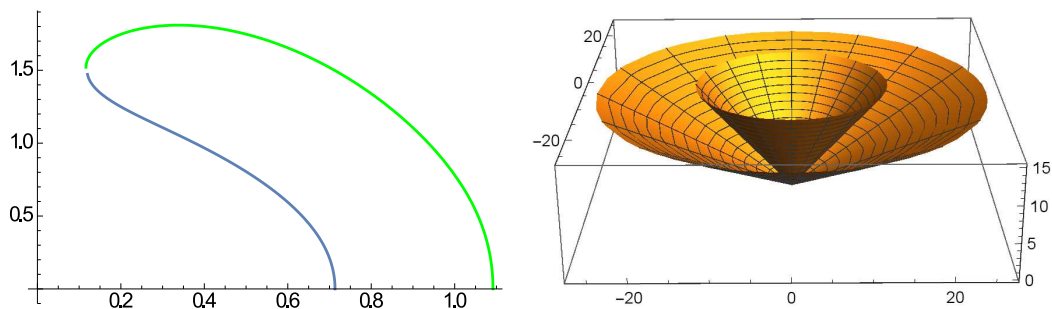} 
\caption{\it On the left: numerical solution of the diff. equation 
\eqref{eom},
obtained by fitting a second piece to the green curve in fig.\ref{num-sol2}. On the right: the inner and outer boundary of the related hollow cone.} 
\label{num-sol3}
\end{figure}

To illustrate the situation by a typical example, we have fine-tuned the numerical
evaluation of the Cauchy initial problem for \eqref{eom} at a point closely located
near the upper endpoint of the green curve in fig.\ref{num-sol2}. The result is 
depicted in fig.\ref{num-sol3}. 
It shows that the smooth continuations of the curves in fig.\ref{num-sol2} 
give us solutions to the boundary problem
\beq
h(\Omega_1)~=~h(\Omega_2)~=~0~,~~~~~\mbox{with}~~\Omega_1~<~\Omega_2~.\label{bc-hollow-cone}
\eeq
With such boundary conditions the volume of our 3-dimensional minimal submanifold \eqref{mf}
via \eqref{holo} is a candidate for the entanglement entropy of special hollow cone regions whose boundaries
are given by the surfaces of two cones with common axis and common tip, but different 
opening angles
$\Omega_1<\Omega_2$. For cones one needs only $\Omega\in(0,\pi/2)$.
To cover all possible interesting cases of hollow cones, one should
keep in mind $\Omega_1\in (0,\pi/2)$ and $\Omega_2\in (0,\pi)$.

The continuation for curves closer and closer  to the original red one in fig.\ref{num-sol2} yields hollow cone solutions with smaller and smaller $\Omega_2-\Omega_1$. 
This shows a remarkable instability of the cone solutions. As soon as one acts on a cone solution with an arbitrary small perturbation away from  fine-tuned Cauchy data, it
immediately jumps to a hollow cone solution with $\Omega_1$ very close to $\Omega_2$.\footnote{As discussed
at the end of this section, for such very thin hollow cones these solutions 
are {\it not} the one with smallest volume and therefore
not responsible for the hollow cone entropy.} 
To stay smoothly within the set of cone solutions, one has to restrict oneself to only fine-tuned
variations of already fine-tuned Cauchy data. 

On the other side, the situation can also be seen as a smooth one. The turns of the 
hollow cone curves in close neighbourhood of the $h$-axis are some kind of smooth
reflections, and in the limit $\Omega_1\rightarrow\Omega_2$ one reaches hard reflection, where one goes back on the infalling curve. 

As a side remark we comment on the situation in the lower dimensional case, which has been mentioned
in the introduction.  There the equation for the analog of $h$ allows generic Cauchy data for all $h>0$.
A small variation away from a certain choice of Cauchy data, which guarantee single cusp boundary conditions,
results in another cusp situation with slightly varied
cusp angle. Hence under variation of Cauchy data these solutions are stable within the class of boundary conditions for a single cusp.\\
 
Let us now turn to the evaluation of the volume of the $AdS$ manifold \eqref{mf},
with $h(\vartheta)$ obeying the hollow cone boundary conditions \eqref{bc-hollow-cone}. At first we have to handle the fact, that now, different from the cone case, 
the relation between $\vartheta$ and $h$ is no longer one to one. Introducing 
some parameter $t$ on the curve in the $(\vartheta,h)$-plot, we define
\beq
h_0~=~\mbox{max}_t~h(t)~,~~~~~\vartheta_0~=~\mbox{min}_t~\vartheta (t)~.\label{h0th0}
\eeq
Both $h_0$ and $\vartheta_0$ are functions of $\Omega_1$ and $\Omega_2$. 

Denoting by
$h_1(\vartheta)\leq h_2(\vartheta)$ the two pieces of the curve $(\vartheta(t),h(t))$, \footnote{The blue and green part in the example of fig.\ref{num-sol3} .} the unregularised expression
for the volume of the $AdS$ submanifold \eqref{mf} with boundary condition \eqref{bc-hollow-cone} becomes
\beq
V^{\mbox{\tiny h.c.}}=2\pi\int _0^{\infty}\frac{d\rho}{\rho}\left (\int_{\vartheta_0}^{\Omega_1} d\vartheta ~\frac{\sin\vartheta}{h_1^3(\vartheta)}\sqrt{1+h_1^2+\dot h_1^2}~+
\int_{\vartheta_0}^{\Omega_2} d\vartheta ~\frac{\sin\vartheta}{h_2^3(\vartheta)}\sqrt{1+h_2^2+\dot h_2^2}\right ).\label{V-hollow}\\[2mm]
\eeq 
For the separation of the divergences it is more convenient to proceed along
the lines of the previous section, changing the integration variable from
$\vartheta$ to $h$. 

On the piece of the curve, described by $h_1(\vartheta)$, the relation between
$\vartheta$ and $h$ is one to one, and we can define $F_1(h)$ similar to \eqref{Fh}
by
\beq
F_1(h)~=~\frac{\sin\vartheta_1(h)}{h^3\dot h_1(\vartheta_1 (h))}\sqrt{1+h^2+\dot h_1^2}~.
\eeq
On the other piece, described by $h_2(\vartheta)$, one has to choose on the parts left and right
of the maximum of $h_2$ the appropriate branch of the (not unique) inversion of the function
$h_2(\vartheta)$. With this in mind we define for $\vartheta\in(\vartheta(h_0),\Omega_2)$
\beq
F_2(h)~=~\frac{\sin\vartheta_2^r(h)}{h^3\dot h_2(\vartheta_2^r(h))}\sqrt{1+h^2+\dot h_2^2}~,~~~h_2(\vartheta_2^r(h))=h, ~~~\frac{d\vartheta_2^r(h)}{dh}<0~,
\eeq
as well as for $\vartheta\in(\vartheta_0,\vartheta(h_0))$
\beq
\hat F_2(h)~=~\frac{\sin\vartheta_2^l(h)}{h^3\dot h_2(\vartheta_2^l(h))}\sqrt{1+h^2+\dot h_2^2}~,~~~h_2(\vartheta_2^l(h))=h, ~~~\frac{d\vartheta_2^l(h)}{dh}>0~.
\eeq

The regularised volume is then
\bea
V^{\mbox{\tiny h.c.}}_{\epsilon,l}&=& ~2\pi\int_{h(\vartheta _0)}^{\epsilon /l} dh ~F_1(h) \int _{\epsilon/h}^{l}\frac{d\rho}{\rho}~+~2\pi\int_{h_0}^{\epsilon /l} dh ~F_2(h) \int _{\epsilon/h}^{l}\frac{d\rho}{\rho}
\label{VF1F2}\\[2mm]
&&~~~~~~~~~~~~~~~~~~~~~~~~ +2\pi\int_{h(\vartheta _0)}^{h_0} dh ~\hat F_2(h) \int _{\epsilon/h}^{l}\frac{d\rho}{\rho}~.\nonumber
\eea
One has $F_2(h)\leq 0$ and $\hat F_2(h)\geq 0$. Near $h_0$ they both become singular like \\$ \mp (h_0-h)^{-1/2}$. Since this singularity is integrable, the third term in \eqref{VF1F2} contributes 
to the divergent piece of $V^{\mbox{\tiny h.c.}}_{\epsilon,l}$ only via
the divergence of the $\rho$-integration. The  
divergences arising from the first two terms are due to divergence of the
$\rho$-integration {\it and} the behaviour of $F_1$ and $F_2$ at $h=0$.
This on its part is determined by $h_1$ and $h_2$ near $\vartheta =\Omega_1$ and $\vartheta =\Omega_2$, respectively. Since both functions are solutions of \eqref{eom} we can apply \eqref{h-expanded}.
The corresponding values for the constant $h_*$ are now tuned differently from the construction
in the previous section. But since anyway its value was not relevant for the evaluation of the
divergent parts we get
\bea
V^{\mbox{\tiny h.c.}}_{\epsilon,l}&=&\frac{\pi}{2}\Big (\sin\Omega_1~+~\sin\Omega_2\Big )\frac{l^2}{\epsilon ^2}~-~\frac{\pi}{8}\Big (\cos\Omega_1~\cot\Omega_1~+~\cos\Omega_2~\cot\Omega_2\Big )~\log ^2\frac{\epsilon}{l}\nonumber\\[2mm]
&&+~\Big (2\pi\int _0^{h(\vartheta_0)}\tilde F_1(h)dh~+~2\pi\int _0^{h_0}\tilde F_2(h)
dh ~+~2\pi\int _{h_0}^{h(\vartheta _0)}\hat F_2(h)dh~\nonumber\\[2mm]
&&~~~~~~~~~~~~~~~~~~~+~\frac{\pi~\sin\Omega_1}{(h(\vartheta_0))^2}~+~\frac{\pi~\cos\Omega_1~\cot\Omega_1}{4}~\log h(\vartheta_0)\label{Vhc}\\[2mm]
&&~~~~~~~~~~~~~~~~~~~+~\frac{\pi~\sin\Omega_2}{h_0^2}~+~\frac{\pi~\cos\Omega_2~\cot\Omega_2}{4}~\log h_0~\Big )~\log\frac{\epsilon}{l}~+~{\cal O}(1)~.\nonumber
\eea
$\tilde F_1$ and $\tilde F_2$ are defined according to \eqref{Ftilde} with the corresponding index. 

The leading and the nextleading divergent terms are just the sum of the corresponding
terms for the two single cones of opening angles $\Omega_1$ and $\Omega_2$. 
For the leading term this is due
to the additivity of the area of $\partial{\cal A}$. For the coefficient
of the  $\log ^2\epsilon$, which is due to the singularities of  $\partial{\cal A}$,
it could have been expected since the complement of a hollow cone consists of two full cones.
Then our calculation shows that touching each other at their tips does not disturbe additivity for the leading and nextleading terms.

The difference
from simple additivity begins with the $\log\frac{\epsilon}{l}$ term. Note that $\vartheta_0$ and $h_0$ are both functions of $\Omega_1$ {\it and} $\Omega_2$.

It is interesting to compare this situation with that for the entanglement
entropy of two regions with just one common point in a lower dimensional setting \cite{Mozaffar:2015xue}. In both cases deviations from additivity appear just
for the  $\log\frac{\epsilon}{l}$ term. However, one has to keep in mind
that $\log\frac{\epsilon}{l}$ terms for $d=3$ appear also in the smooth case, while
for $d=2$ they are special for regions with singularities in their boundary.\\

Of course the disconnected 3-dimensional manifold consisting out of the manifolds for the
two single cones \footnote{To be precise: disconnected inside $AdS$, the two pieces
touch each other on the boundary of $AdS$ at the common tip of the cones.} always fulfils also our differential equation \eqref{eom} and the hollow cone
boundary conditions \eqref{bc-hollow-cone}. The decision for which angles $\Omega_1,\Omega_2$
the connected version, by its smaller volume, is favoured with respect to the disconnected
version needs involved numerical analysis of the $\log\frac{\epsilon}{l}$ term of both
candidates.

\begin{figure}[h!]
 \centering
 \includegraphics[width=11cm]{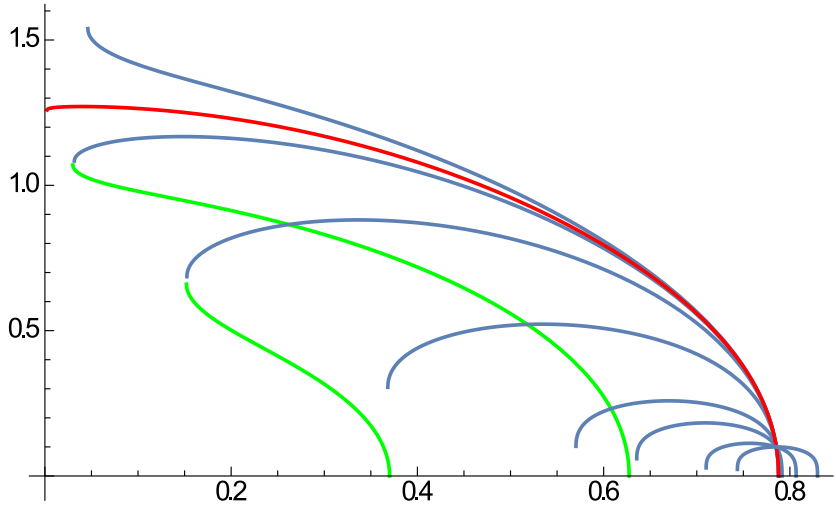} 
\caption{\it Numerical solutions $h(\vartheta )$ of \eqref{eom} with $h(\pi/4)=0.1$ and $\dot h(\pi/4)\in (0,-1,-5,-10,-18,-19.5,-19.7,-19.724,-19.75)$, plotted
for $h>0$ and down to values of $\vartheta$, where $\dot h$ diverges (in blue, in red for the approximate cone solution). For two of these curves approximate smooth continuations are shown in green.}
\label{all-hollow}
\end{figure}

However, one gets a clear qualitative picture by plotting solutions obtained
from Cauchy data at a point very close to the $\vartheta$-axis, see fig.\ref{all-hollow}.
For small initial values of $\vert\dot h\vert$ the curve stays in the vicinity of the initial point.
 Choosing negative $\dot h $, but larger  $\vert\dot h\vert$, the curves get larger
maxima $h_0$, their minima in $\vartheta$, i.e. $\vartheta_0$, move further down on the $\vartheta$-axis
and the angular thickness of the related hollow cone, the difference $\Omega_2-\Omega_1$, becomes larger.
One can continue until $\vartheta _0$ approaches zero. But remarkably, somewhere in between the angular thickness $\Omega_2-\Omega_1$ no longer increases and  goes back to zero instead. 

From this observation we learn the following lessons.  To each outer angle $\Omega_2$
belongs a limiting value $\Omega_{\mbox{\tiny min}}(\Omega_2)$ for $\Omega_1$.  For $\Omega_{\mbox{\tiny min}}(\Omega_2)<\Omega_1<\Omega_2$ one gets two connected solutions. 
For hollow cones with very small $\Omega_2-\Omega_1$ the one with smaller $h_0$ has smaller volume. 
This can be seen as following. The coefficients of the two leading divergences in \eqref{Vhc} are the same for both solutions. The coefficient of the logarithmic divergence stands in front of a negative term and goes to
plus infinity for $h_0,h(\vartheta _0)\rightarrow 0$. 
Obviously, by an analogous reasoning, the volume of this favoured connected solution is also smaller than that of the corresponding disconnected solution. 

Furthermore, there exists no connected solution if $0<\Omega _1<\Omega_{\mbox{\tiny min}}(\Omega_2)$. Hence in varying 
$\Omega_1$ from values near  $\Omega_2$  to such below $\Omega_{\mbox{\tiny min}}(\Omega_2)$ , there has to be a transition
of the related entanglement entropy from connected to disconnected solutions. 
Whether it happens at once from the connected solution with the smaller $h_0$
to the disconnected solution, or whether there appears
first a transition between the two connected solutions, has to be left open for
further study. These should also clarify the order of that geometrically induced phase transition pattern.

\begin{figure}[h!]
 \centering
 \includegraphics[width=14cm]{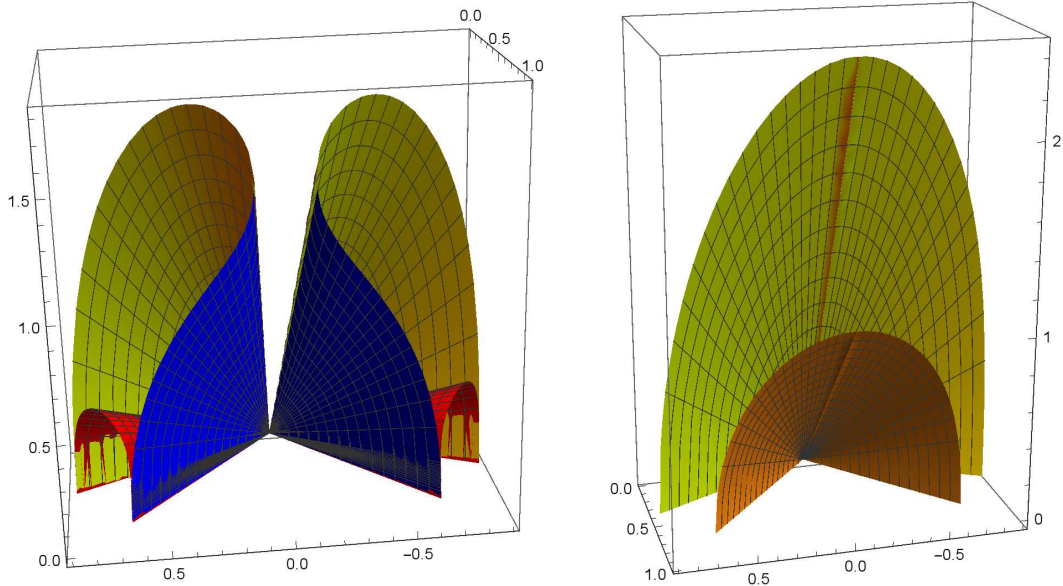} 
\caption{\it The bottom is the subspace $x_2=0$ of the boundary of $AdS_4$. The upward direction points inside  $AdS$. On the left: the two connected solutions. On the right: the disconnected solution. The corresponding hollow cone is that from fig.\ref{num-sol3}.} 
\label{AdS}
\end{figure}

We close this section with a visualisation of the extension of the connected 
and disconnected hollow cone solutions into the interior of $AdS$. Fig.\ref{AdS}  shows the intersection of the 3-dimensional submanifolds corresponding to the hollow cone of fig.\ref{num-sol3} with the codimension one subspace $x_2=0$.    
One
should not be confused by the fact that the projections shown in the left part look disconnected, the full 3-dimensional manifolds are connected via the suppressed $x_2$-direction.
\section{Elementary geometry of banana shaped regions}
As an example for a compact region, whose boundary has two conical singularities, we take
the banana shaped region  obtained by a suitable conformal map of the infinite cone. We
apply the inversion at the unit sphere
\beq
x_{\mu}~\mapsto ~\frac{x_{\mu}}{x^2} \label{inversion}
\eeq
to an infinite cone of opening angle $2\Omega$ with its tip located at $x_{\mu}=(0,0,q)$
and its axis being situated in the $(x_1,x_3)$-plane with an angle $\alpha$ relative to the $x_3$-axis. Obviously it is sufficient to restrict both $\Omega$ and $\alpha$
to the interval $(0,\pi/2)$.  

\begin{figure}[h!]
 \centering
 \includegraphics[width=11cm]{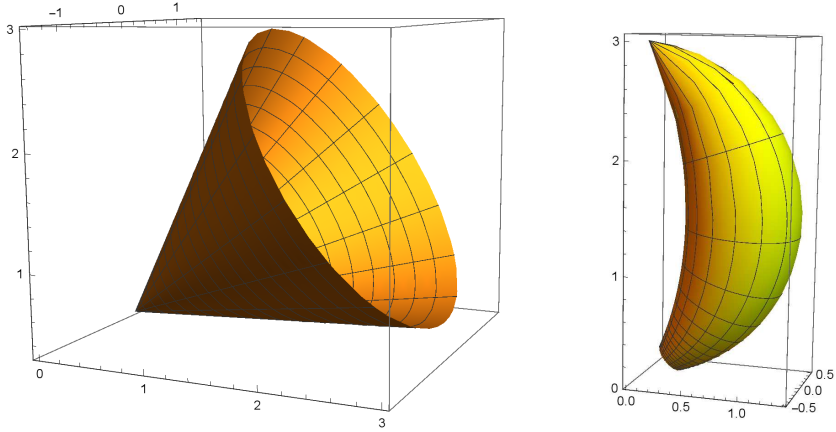} 
\caption{\it The map \eqref{inversion} for $q=1/3,~\alpha =1,~\Omega =0.5$. Both for the preimage and 
the $~~~~~~~~~~~~~~$image we show $0<\rho <3$, only. } 
\label{banana-plot}
\end{figure}

This procedure gives the region, see fig.\ref{banana-plot},
\bea
x_1(\rho,\vartheta,\varphi )&=&\frac{\rho ~\cos\alpha~\sin\vartheta~\cos\varphi~~+~\rho~\sin\alpha~\cos\vartheta}{q^2+\rho ^2~+~2q\rho ~(\cos\alpha ~\cos\vartheta-\sin\alpha ~\sin\vartheta~\cos \varphi )} ~,\nonumber\\[2mm]
x_2(\rho,\vartheta,\varphi )&=&\frac{\rho~\sin\vartheta~\sin\varphi}{q^2+\rho ^2~+~2q\rho ~(\cos\alpha ~\cos\vartheta-\sin\alpha ~\sin\vartheta~\cos \varphi )} ~,\nonumber \\[2mm]
x_3(\rho,\vartheta,\varphi )&=&\frac{q~+~\rho~\cos\alpha~\cos\vartheta~-~\rho~\sin\alpha~\sin\vartheta~\cos\varphi}{q^2+\rho ^2~+~2q\rho ~(\cos\alpha ~\cos\vartheta-\sin\alpha ~\sin\vartheta~\cos \varphi )}~.\label{banana}
\eea
The coordinates $\rho,\vartheta$ and $\varphi $ obey
\beq
0~\leq~\rho ~<\infty ~,~~~~0~\leq ~\vartheta ~\leq ~\Omega~,~~~0~\leq ~\varphi ~< ~2\pi~.
\eeq

The transformation \eqref{inversion} preserves angles and maps spheres to spheres and circles to circles (planes and straight lines understood as spheres and circles passing infinity). 

Therefore, the
boundary surface of the region \eqref{banana} has two conical singularities of opening angle $2\Omega$, whose distance $D$ is
\beq
D~=~\frac{1}{q}\label{Dq}
\eeq
and which are located at $(0,0,1/q)$ and at the origin. This boundary surface is described by
 $x_{\mu}(\rho,\Omega,\varphi)$, with $\rho $ and $\varphi$ as coordinates. The lines of constant $\varphi $ are segments of circles passing the two singular points and having the radii
\beq
R_{\alpha,\Omega,q}(\varphi)~=~\frac{1}{2q\sqrt{1-(\cos\alpha~\cos\Omega-\sin\alpha~\sin\Omega~\cos\varphi)^2}}~.
\eeq
The minimal and maximal values of these radii are
\beq
R_{\alpha,\Omega,q}^{~\mbox{\tiny min/max}}~=~\frac{1}{2q~\vert\sin (\alpha \pm \Omega )\vert }~.\label{Rminmax}
\eeq 
The axis of the banana \eqref{banana} is a segment of a circle of radius
\beq
R_{\alpha,\Omega,q}^{~\mbox{\tiny axis}} ~=~\frac{1}{2q~\sin\alpha}~.\label{Raxis}
\eeq

Furthermore, the lines of constant $\rho$ are complete circles. They  are
intersections of the boundary surface with two-spheres, touching one of the singular points and having their center in the plane dividing the banana into two halves. Among those circles the largest one has a radius
\beq
 R_{\alpha,\Omega,q}^{\perp}~=~\frac{\sin\Omega}{2q~(\cos\alpha+\cos\Omega )}      ~.
\label{Rperp}
\eeq

The metrical geometry of the boundary surface of our banana shaped region \eqref{banana} is fixed by the dimensionful parameter $q$ and the two angles $\alpha$ and $\Omega $. Using (\ref{Dq},\ref{Rminmax},\ref{Raxis},\ref{Rperp}) these parameters can be expressed in terms of three out of the parameters measuring length distances 
related to the boundary surface.\\ 

For later use we are still interested in the area of the boundary surface. The square
root of the induced metrics determinant is
\beq
\sqrt{g}~=~\frac{\rho ~\sin\Omega}{\big (q^2+\rho ^2+2q\rho ~(\cos\alpha~\cos\Omega -\sin\alpha~\sin\Omega~\cos\varphi )\big )^2}~.
\eeq
Performing either the $\varphi $-integration or the $\rho$-integration we arrive at \footnote{The remaining integrals can be expressed in terms
of elliptic functions, but we did not find a short compact expression.}
\bea
A_{\alpha,\Omega,q}&=&\frac{2\pi ~\sin\Omega}{q^2} \int _0^{\infty}\frac{x ~ (1+x^2+2x~\cos\alpha~\cos\Omega )~ d
x}{\big ( (1+x ^2+2x~\cos (\alpha +\Omega ))(1+x^2+2x~\cos (\alpha -\Omega ))\big )^{3/2}}~,\nonumber\\[2mm]
&=&\frac{\sin\Omega}{2q^2} \left (\frac{\pi}{\cos\alpha +\cos\Omega}+\frac{\pi}{\vert \cos\alpha -\cos\Omega \vert}-\int _0^{2\pi} \frac{\hat w (\varphi )~\mbox{arccos}~\hat w}{(1-\hat w^2)^{3/2}} ~d\varphi \right )~. \label{area1}
\eea
In the second line also part of the $\varphi$-integration has been performed using the integral
\beq
\int _0^{2\pi}\frac{d\varphi }{1-\hat w^2(\varphi)}~=~\frac{\pi}{\cos\alpha +\cos\Omega}~+~\frac{\pi}{\vert \cos\alpha -\cos\Omega \vert}~,\label{1-wsquared}
\eeq
where
\beq
\hat w(\varphi )~=~\cos\alpha~\cos\Omega ~-~\sin\alpha ~\sin\Omega ~\cos\varphi ~.
\label{what}
\eeq
\\ 
Some special cases of \eqref{area1} are
\bea
A_{\Omega,\Omega,q}&=&\frac{\pi}{16 q^2}~\frac{2~\sin\Omega+2~\sin ^3\Omega-\cos ^4\Omega~\log\big (\frac{1+\mbox{\small sin}\Omega}{1-\mbox{\small sin}\Omega}\big )}{\sin ^2\Omega~\cos ^2\Omega}~,\nonumber\\[2mm]
A_{0,\Omega,q}&=&\frac{\pi}{q^2~\sin\Omega }~\big (1-\Omega ~\cot\Omega \big )~,\nonumber\\[2mm]
A_{\alpha,\pi /2,q}&=&\frac{\pi }{q^2~\cos ^2\alpha}~.\label{area-special}
\eea

\section{Entanglement entropy for banana shaped regions}
As in the previous section we shift the tip of the cone in the boundary of $AdS$
out of the origin and tilt its axis relative to the $x_3$-axis. Then the corresponding 3-manifold in $AdS_4$ is
\bea
x_1&=&\rho ~\cos\alpha~\sin\vartheta~\cos\varphi~~+~\rho~\sin\alpha~\cos\vartheta ~,\nonumber\\
x_2&=&\rho~\sin\vartheta~\sin\varphi ~,\nonumber \\
x_3&=&q~+~\rho~\cos\alpha~\cos\vartheta~-~\rho~\sin\alpha~\sin\vartheta~\cos\varphi~,\nonumber\\
r&=&\rho~h(\vartheta)~,\label{preimage}
\eea
with $h(\vartheta )$ the same function as in section 2.

Now we apply the mapping
\beq
x_{\mu}~\mapsto ~\frac{x_{\mu}}{x^2+r^2}~,~~~~r~\mapsto ~\frac{r}{x^2+r^2}~,\label{map}
\eeq
which is an $AdS$-isometry and acts conformally on the boundary $r=0$. Then
we get the submanifold of our interest
\bea
x_1&=&\frac{\rho ~(\cos\alpha~\sin\vartheta~\cos\varphi~+~\sin\alpha~\cos\vartheta)}{q^2~+~\rho^2(1+h^2(\vartheta)) ~+2q \rho ~w(\vartheta,\varphi)},\nonumber\\[2mm]
x_2&=&\frac{\rho~\sin\vartheta~\sin\varphi}{q^2~+~\rho^2(1+h^2(\vartheta)) ~+2q \rho ~w(\vartheta,\varphi)},\nonumber\\[2mm]
x_3&=&\frac{q~+~\rho~w(\alpha,\vartheta,\varphi)}{q^2~+~\rho^2(1+h^2(\vartheta)) ~+2q \rho ~w(\vartheta,\varphi)},\nonumber\\[2mm]
r&=&\frac{\rho ~h(\vartheta)}{q^2~+~\rho^2(1+h^2(\vartheta)) ~+2q \rho ~w(\vartheta,\varphi)}~,\label{surface}
\eea
with
\beq
w(\vartheta,\varphi)~=~\cos\alpha~\cos\vartheta~-~\sin\alpha~\sin\vartheta~\cos\varphi~.\label{w}
\eeq
Here $\alpha$ and $q$ are parameters specifying the submanifold, and
$\rho,~\vartheta,~\varphi$ are coordinates.

Since the map \eqref{map} is an isometry, the volume of the manifold \eqref{surface} is given by an integral over the coordinates with the same integrand as in section 2. But now there is no need for an IR-regularisation and
the cut-off for the UV-regularisation introduces modified boundaries for the coordinates.

The condition $r=\epsilon$
has two solutions for $\rho$ as a function of $h$ and $\varphi$
\beq
\rho^{(\pm)}_{\epsilon}(h,\varphi )~=~\frac{h-2q\epsilon~w(\vartheta (h),\varphi)\pm\sqrt{(h-2q\epsilon w)^2-4\epsilon ^2q^2(1+h^2)}}{2\epsilon~(1+h^2)}~.\label{rhopm}
\eeq
Let us introduce for later convenience
\beq
N_{\epsilon}(h,\varphi)~=~1-\frac{2q\epsilon}{h}w(\vartheta(h),\varphi)~+~\sqrt{\big (1-\frac{2q\epsilon w}{h}\big )^2-4\epsilon^2q^2\frac{1+h^2}{h^2}}~.\label{N}
\eeq
Then one has
\beq
\rho^{(+)}_{\epsilon}~=~\frac{h}{2\epsilon(1+h^2)}\cdot N_{\epsilon}(h,\varphi)~,~~~~
\rho^{(-)}_{\epsilon}~=~\frac{2q^2\epsilon}{h}\cdot\frac{1}{N_{\epsilon}(h,\varphi)}~,\label{rhopm2}
\eeq
and the interval for the $\rho$-integration will be
\beq
\rho^{(-)}_{\epsilon}~<~\rho ~<~\rho^{(+)}_{\epsilon}~.\label{rhointerval}
\eeq
The expression under the square root in \eqref{rhopm} has to be positive. This 
constrains $h$ from below, i.e.
\beq
h_{\epsilon}~<~h~<h_0~,\label{h-interval}
\eeq
where $h_0$ as a function of $\Omega$ has been defined in section 2
and  
\beq
h_{\epsilon}(\varphi )~=~\frac{2q\epsilon}{1-4q^2\epsilon ^2}~\Big ( w(\vartheta(h_{\epsilon}),\varphi)~+~\sqrt{1-4q^2\epsilon ^2+4q^2\epsilon ^2w^2}~\Big )~.\label{heps}
\eeq 
Note that $h_{\epsilon}$ appears also on the r.h.s. as argument of the function $\vartheta(h)$. The reality condition for the square root in \eqref{rhopm} allows
also a minus in front of the square root in \eqref{heps}, but the analysis for
small $\epsilon$ shows, that one has to choose the positive sign to ensure
positive values for $h_{\epsilon}$. With \eqref{sintheta} we find the expansion
\bea
h_{\epsilon}(\varphi )&=&B_1~q\epsilon ~+~B_3~q^3\epsilon ^3~+~\dots ~,~~~~\mbox{with}\label{heps-expanded}
\\
B_1(\varphi)&=&2\big (1~+~\hat w(\varphi)\big )~,\label{B1}\\
B_3(\varphi)&=&2\big (1~+~\hat w(\varphi)\big )^2~\big (2+\cos\Omega(\cos\alpha+\sin\alpha~\cot\Omega~\cos\varphi)\big )~.\label{B3}
\eea
The function $\hat w(\varphi)=w(\Omega,\varphi)$
has been defined in \eqref{what}, see also \eqref{w}.

After this discussion of the region of integration, the regularised volume is

\beq
V_{\epsilon}~=~-\int_0^{2\pi}d\varphi\int_{h_{\epsilon}}^{h_0}dh\int_{\rho^{(-)}_{\epsilon}}^{\rho^{(+)}_{\epsilon}}\frac{d\rho}{\rho}~F(h)~.
\eeq
Performing the trivial $\rho$ integration we get with \eqref{rhopm2}
\beq
V_{\epsilon}~=~V_{\epsilon}^{(1)}~+~V_{\epsilon}^{(2)}~+~V_{\epsilon}^{(3)}~,
\eeq
where
\bea
V_{\epsilon}^{(1)}&=&-2~\int_0^{2\pi}d\varphi\int_{h_{\epsilon}(\varphi)}^{h_0}dh~ F(h)~\log N_{\epsilon}(h,\varphi)~,\nonumber\\[2mm]
V_{\epsilon}^{(2)}&=&2~\log(2q\epsilon)~\int_0^{2\pi}d\varphi\int_{h_{\epsilon}(\varphi)}^{h_0}dh~ F(h)~,\nonumber\\[2mm]
V_{\epsilon}^{(3)}&=&\int_0^{2\pi}d\varphi\int_{h_{\epsilon}(\varphi)}^{h_0}dh~ F(h)~\log 
\big (\frac{1+h^2}{h^2}\big )~.
\eea
The function $F(h)$ has been defined in \eqref{Fh}. To handle the divergence of $F(h)$ at $h\rightarrow 0$ we
use \eqref{Ftilde} with $\tilde F~=~{\cal O}(h)$. 

Now $V_{\epsilon}^{(1)}$  is the most involved part, since, besides the manifest $h$-dependence of its integrand, there is also one via $\vartheta(h)$ in $N_{\epsilon}$, see \eqref{N}. Therefore let us start with
\beq
V_{\epsilon}^{(2)}~=~2~\log(2q\epsilon)~\int_0^{2\pi}d\varphi\int_{h_{\epsilon}(\varphi)}^{h_0}dh~ \Big (\tilde F(h)~+~\frac{\cos\Omega ~\cot\Omega}{8h}~-~\frac{\sin\Omega}{h^3}  \Big )~.
\eeq
Up to terms vanishing for $\epsilon\rightarrow 0$ we can replace in the integral
over $\tilde F(h)$ the lower boundary by zero. The integrations over $1/h^3$ and $1/h$ are trivial, and with \eqref{heps-expanded} we get altogether
\bea
V_{\epsilon}^{(2)}&=&-\Big (\frac{\log(q\epsilon)}{q^2\epsilon ^2}~+~\frac{\log 2}{q^2\epsilon ^2}~\Big )~\sin\Omega\int_0^{2\pi}\frac{d\varphi}{B_1^2}~-~\frac{\pi}{2}\cos\Omega~\cot\Omega~\log ^2(q\epsilon )\label{V2final}\\
&&+~\big (\log (q\epsilon )~+~\log 2\big )\left (4\pi\int _0^{h_0}\tilde Fdh~+~2~\sin\Omega~\Big (\frac{\pi}{h_0^2}~+~\int_0^{2\pi}\frac{B_3}{B_1^3}~d\varphi \Big ) \right .\nonumber\\
&&\left .~~~~~~~~~~~~~~~~~~~~~~~~~~~~+\cos\Omega~\cot\Omega~\Big (\frac{\pi}{2}~\log h_0~-~\frac{1}{4}\int_0^{2\pi}\log B_1d\varphi \Big )\right )\nonumber\\
&&
-~\frac{\pi~\log 2}{2}\cos\Omega~\cos\Omega ~\log(q\epsilon)+o(1)~.\nonumber
\eea

A similar treatment yields
\bea
V_{\epsilon}^{(3)}&=&\frac{\log(q\epsilon)}{q^2\epsilon ^2}~\sin\Omega\int_0^{2\pi}\frac{d\varphi}{B_1^2}~+~\frac{\sin\Omega}{q^2\epsilon ^2}\int_0^{2\pi}\Big (\frac{1}{2B_1^2}+\frac{\log B_1}{B_1^2}\Big )d\varphi\label{V3final}\\[2mm]
&&+~\frac{\pi}{4} \cos\Omega~\cot\Omega~\log ^2 (q\epsilon )\nonumber\\
&&+~\log (q\epsilon)~\left (2\sin\Omega\Big (\pi-\int_0^{2\pi}\frac{B_3}{B_1^3}~
d\varphi \Big )~+~\frac{1}{4}~\cos\Omega~\cot\Omega~\int_0^{2\pi}\log B_1~d\varphi \right )\nonumber
\\
&&+~2\pi\int_0^{h_0}\tilde F~\log\frac{1+h^2}{h^2}~dh\nonumber\\
&&+~\sin\Omega~\left (\int_0^{2\pi}\Big (1-2\frac{B_3}{B_1^3}\Big )~\log B_1~d\varphi +\pi\frac{1+h_0^2}{h_0^2}~\Big (\log\frac{1+h_0^2}{h_0^2}-1\Big )\right )\nonumber\\
\nonumber\\
&&+~\cos\Omega~\cot\Omega ~\left ( \frac{1}{8}\int _0^{2\pi}\log ^2B_1~d\varphi -\frac{\pi}{8}~\mbox{Li}_2(-h_0^2)-\frac{\pi}{4}~ \log ^2h_0 \right )~+~o(1)~.\nonumber
\eea
  
To start the discussion of the asymptotics of $V_{\epsilon}^{(1)}$ at $\epsilon\rightarrow 0$, we note
that at any fixed $0\leq\varphi\leq 2\pi$, $h>0$
\beq
N_{\epsilon}(h,\varphi)~=~2~+~{\cal O}(\epsilon)~.
\eeq
Since $\tilde F$ tends to zero linearly in $h$, this estimate can be used even if the
lower boundary of the $h$-integration tends to zero. This implies
\beq
V_{\epsilon}^{(1)}~=~-4\pi~\log 2~\int_0^{h_0}\tilde F(h)dh~+~\sin\Omega~J_1~+~\cos\Omega~\cot\Omega~J_2~+~o(1)~,\label{V1}
\eeq
with
\beq 
J_1~=~2\int _0^{2\pi}d\varphi\int_{h_\epsilon(\varphi)}^{h_0}\frac{\log N_{\epsilon}(h,\varphi )}{h^3}dh~,~~~J_2~=-~\frac{1}{4}\int _0^{2\pi}d\varphi\int_{h_\epsilon(\varphi)}^{h_0}\frac{\log N_{\epsilon}(h,\varphi )}{h}dh~.\label{J1J2}
\eeq

The estimate of these two integrals is performed in appendix A. 
Putting \eqref{J1J11J12} and the results
\eqref{J11final},\eqref{J12final},\eqref{J2final}  into \eqref{V1} we get
\bea
V_{\epsilon}^{(1)}&=&\frac{2~\sin\Omega}{q^2\epsilon^2}\int _0^{2\pi}d\varphi ~
\Big (\frac{1}{8(1-\hat w^2)}-\frac{1}{16(1+\hat w)^2}
-\frac{\hat w~\mbox{arccos}~\hat w}{8(1-\hat w^2)^{3/2}}-\frac{\log (1+\hat w)}{8(1+\hat w)^2} \Big )
\nonumber\\[2mm]
&&+~\frac{\pi\log 2}{2}~\cos\Omega~\cot\Omega~\log(q\epsilon )\nonumber\\
&&-~\sin\Omega~\left ( 4\pi+\frac{2\pi\log2}{h_0^2}+4\int _0^{2\pi}d\varphi\int_0^{1/B_1}\frac{K(\varphi )}{\sqrt{(1-2 x\hat w)^2-4x^2}}~dx \right .\nonumber\\
&&~~~~~~~~~~~~~~~~~~~~~~~~~~~~~~~~~~~~~~~~~~~~~\left . -2\int _0^{2\pi}d\varphi~\frac{B_3}{B_1^3}~\log (1+\hat w(\varphi )) \right )\nonumber\\[2mm]
&&+~\cos\Omega~\cot\Omega~\left (~\frac{\log 2}{4}\int _0^{2\pi}\log B_1d\varphi~-~\frac{\pi \log2}{2}~\log h_0  \right. \nonumber\\ 
&&\left. ~~~~
-\frac{1}{4}\int _0^{2\pi}d\varphi\int _0^{1/B_1}\log \Big (\frac{1}{2}-x\hat w+\frac{1}{2}\sqrt{\big (1-2x\hat w)^2-4x^2}~\Big )\frac{dx}{x} ~\right ) \nonumber\\[2mm]
&&-~4\pi~\log2~\int _0^{h_0}\tilde F(h)dh~+~o(1)~.\label{V1final}
\eea
Remember that $B_1$, $B_3$ and $\hat w$ as functions of $\varphi$ are defined in \eqref{B1} and \eqref{what}. Furthermore, from \eqref{K} in appendix A we know
\beq
K(\varphi )~=~\frac{1}{2}~\Big (\frac{B_3}{B_1^2}-1\Big )~=~\frac{1}{4}~\big (\cos\alpha~\cos\Omega+\sin\alpha~\cos\Omega~\cot\Omega~\cos\varphi \big )~.\label{K-text}
\eeq
\\
Now  we have to add \eqref{V1final},\eqref{V2final} and \eqref{V3final} and get
\beq
V_{\epsilon}~=~\frac{P_2}{q^2 \epsilon ^2}~+~L_2~\log ^2(q\epsilon)~+~L_1~\log(q\epsilon)~+~V_0~+~o(1)~.\label{Vfinal}
\eeq
In comparison with the general structure \eqref{odd-GW}, mentioned in the introduction, this means $c_1=P_2/q^2$, $a=L_1$ and $c_3=V_0$.
Using also \eqref{1-wsquared} and
\beq
\int _0^{2\pi}\log B_1(\varphi )d\varphi ~=~2\pi ~\log\big ( (1+\cos\alpha)(1+\cos\Omega )\big )~,
\eeq
the coefficients in \eqref{Vfinal} are given by
\bea
P_2&=&\frac{\sin\Omega}{4}~ \left (\frac{\pi}{\cos\alpha +\cos\Omega}+\frac{\pi}{\vert \cos\alpha -\cos\Omega \vert}-\int _0^{2\pi} \frac{\hat w (\varphi )~\mbox{arccos}~\hat w}{(1-\hat w^2)^{3/2}} ~d\varphi \right )~, \label{P2}\\[2mm]
L_2&=&-~\frac{\pi}{4}~\cos\Omega~\cot\Omega~,\label{L2}\\[2mm]
L_1&=&4\pi\int_0^{h_0}\tilde F(h)~dh~+2\pi\big (1+\frac{1}{h_0^2}\big )~\sin\Omega~+~
\frac{\pi}{2}~\cos\Omega~\cot\Omega~\log h_0~
\eea
and the finite part by
\bea
V_0&=&\sin\Omega ~\left (2\pi~\log\big ((1+\cos\alpha )(1+\cos\Omega )\big) +\frac{\pi (1+h_0^2)}{h_0^2}\Big (\log\frac{1+h_0^2}{h_0^2}-1\Big )\right .\label{V0}\\
&& ~~~~~~~~~~~~~~~~~~~~~~~~-4\pi\left . -4\int _0^{2\pi}\int _0^{1/B_1}\frac{K(\varphi )}{\sqrt{(1-2 x\hat w(\varphi ))^2-4x^2}}~dx \right )\nonumber\\[2mm]
&&+~\cos\Omega~\cot\Omega~\left (\frac{1}{8}\int _0^{2\pi}\log ^2 B_1(\varphi )~d\varphi- \frac{\pi}{8}\mbox{Li}_2(-h_0^2)-\frac{\pi ~\log^2h_0}{4} \right .\nonumber \\[2mm]
&&~~~~-\left .\frac{1}{4}\int _0^{2\pi}d\varphi\int _0^{1/B_1}\log \Big (\frac{1}{2}-x\hat w(\varphi )+\frac{1}{2}\sqrt{\big (1-2x\hat w)^2-4x^2}~\Big )~\frac{dx}{x}~\right )\nonumber\\
&&+~2\pi ~\int _0^{h_0}\tilde F(h)~\log\frac{1+h^2}{h^2}~dh ~.\nonumber
\eea
For clarity let us repeat the meaning of the entries in these formulae. $\Omega,\alpha,q$ are geometrical parameters of the banana
shaped region. In detail, $2\Omega$ is the opening angle of its tips, $\alpha$ is the angle between its axis and the straight line
connecting the tips, $1/q$ is the distance between the tips. $h_0$ is a function of $\Omega$ and determined by the cone
solution \eqref{mf} as the maximal value of $r/\rho$, realised at the cone axis. The technical functions $\hat w(\varphi ),B_1(\varphi)$ and $K(\varphi)$ are defined in eqs. \eqref{what},\eqref{B1} and \eqref{K-text}.
\\

Comparing \eqref{P2} and \eqref{area1}
we find as expected 
\beq
\frac{P_2}{q^2}~=~\frac{A_{\alpha,\Omega,q}}{2}~. 
\eeq

The straightforward application of a similar conformal transformation to the hollow cones of section 3, would
give us the holographic entanglement entropy of banana shaped regions with a more narrow one cutted out. 
\section{Conclusions}
The main result of this paper is the explicit calculation of the regularised volume
of the minimal submanifold in $AdS$, needed for the holographic entanglement entropy of an exemplary {\it compact} three-dimensional
region ${\cal A}$, whose boundary has two conical singularities. 
As for regions with smooth boundary, the coefficient of the $1/\epsilon^2$ term is equal to half of the area of the boundary $\partial {\cal A}$.

The nextleading divergent term is due to the two conical singularities of our
$\partial {\cal A}$ and as expected of squared logarithm type.
Its coefficient is twice the coefficient obtained for an infinite cone with the same opening angle.

Divergences proportional to $\log\epsilon$ are present also for smooth
boundaries and have conformally invariant coefficients. This
invariance holds also in our case since 
the corresponding coefficient depends on the opening angle only.

There is however a subtlety in comparison with the {\it infinite} cone, which
can be obtained as the image of our {\it compact} region ${\cal A}$ under an inversion on a
sphere, whose center is located at one of the singular points of $\partial {\cal A}$.
Under such exceptional conformal transformations one finds  an anomaly also
for smooth boundaries, as discussed in appendix B. Here we observe, that the
coefficient for the banana shaped ${\cal A}$ is twice that for the cone plus
the term $2\pi~\sin\Omega$. The factor 2 arises somehow naturally by coupling
in the cone cases UV and IR regularisation by $l=1/\epsilon$. The additive anomaly 
term approaches for $\Omega\rightarrow \pi/2$ the
anomaly for smooth boundaries as exemplified for spheres versus planes in 
appendix B. 

For generic ${\cal A}$ in $d=3$ the finite piece of the regularised volume
is not expected to be conformally invariant. Of course we observe this also
in our banana case. It depends on $\Omega$ and the angle $\alpha$ between
the axis of the banana and the straight line through its two conical tips.  
While $\Omega$ is invariant under conformal maps, $\alpha$ is not.\\

A second set of observations arose in connection with the following issue.
Solutions of the nonlinear second order differential equation for $h(\vartheta )$, 
governing the minimal submanifold for the case where ${\cal A}$ is a cone with 
opening angle $2\Omega$, have to satisfy the boundary condition $h(\Omega )=0$ {\it and} the requirement $h(0)>0$. Generation of such solutions via
a Cauchy initial value problem inside $(0,\Omega)$ requires suitable
fine-tuning between $h$ and $\dot h$. In the analogous case in $d=2$, a small
perturbation of a given fine-tuned choice of Cauchy data results in a small
variation of the cusp boundary conditions. In our three-dimensional case
perturbing the fine-tuning,  results immediately
in a solution satisfying $h(\Omega_1)=h(\Omega_2)=0$ for some $\Omega_1<\Omega_2$
close to $\Omega$. But these are then boundary conditions for hollow cones.

We have calculated the divergent pieces of the regularised volume relevant
for the holographic entanglement entropy for regions ${\cal A}$ chosen
as a hollow cone, parameterised by the opening angles of its inner and outer
boundary. The coefficients of the $1/\epsilon^2$ term and the $\log ^2\epsilon$
are simply equal to the sums of the corresponding terms for two full cones with 
angle $\Omega_1$ and $\Omega_2$, respectively. 
Nontrivial dependence on both angles  
starts with the coefficient of $\log\epsilon$.

Depending on the two angles, one has to check whether a connected submanifold,
as studied in section 3, or the disconnected one, whose two pieces each correspond to
a single full cone, are favoured by its smaller volume. We have shown that to each
$\Omega_2$ belongs  a bound for $\Omega_1$, such that below this bound only
the disconnected solution exists. Above this bound one finds two connected solutions.
Among them, at least for very small $\Omega_2-\Omega_1$, the one which stays closer to the boundary of $AdS$ is favoured.
It would be interesting to fully explore the pattern of this type of geometrically induced phase transition,
as well as its embedding in a setting with temperature. 
In connection with such an analysis also the subadditive inequality for the entanglement entropy could be of interest.
\footnote{For an application to the cusp case in $d=2$ see \cite{Hirata:2006jx,Hirata:2008ms}.}    

Since the complement of a hollow cone consists of two regions touching each other only at the tip of the hollow
cone,  our observed  angle dependent phase transition resembles the distance dependent transitions
of the entanglement entropy for disconnected regions discussed in \cite{Headrick:2010zt,Tonni:2010pv,Mozaffar:2015xue}.  

\vspace*{10mm}
 \noindent
 {\bf Acknowledgement:}\\[2mm]
I would like to thank Danilo Diaz for earlier discussions on entanglement entropy and George Jorjadze for numerous useful discussions over the course of this work.\\[10mm]
\section*{Appendix A}
Here we discuss in detail the evaluation of $J_1$ and $J_2$, defined in \eqref{J1J2}, for $\epsilon\rightarrow 0$. With the substitution $x=q\epsilon /h$ and
\beq
x_{\epsilon}(\varphi )~=~\frac{1}{B_1(\varphi )}~-~q^2\epsilon ^2\frac{B_3(\varphi )}{B_1^2(\varphi )}~+~\dots \label{xeps}
\eeq
we get
\beq
J_1~=~\frac{2}{q^2\epsilon ^2}~\int _0^{2\pi}d\varphi\int _{q\epsilon /h_0}^{x_{\epsilon}} x~\log \Big (1-2xw(\theta,\varphi)+\sqrt{(1-2xw)^2-4x^2-4q^2\epsilon ^2}~\Big ) dx
\eeq
and after expanding the integrand in $\epsilon $
\beq
J_1~=~J_{11}~+~J_{12}~+~o(1)~,\label{J1J11J12}
\eeq
with
\bea
J_{11}&=&\frac{2}{q^2\epsilon ^2}~\int _0^{2\pi}d\varphi\int _{q\epsilon /h_0}^{x_{\epsilon}} x~\log \Big (1-2xw(\vartheta,\varphi)+\sqrt{(1-2xw)^2-4x^2}~\Big ) dx~,  \label{J11}\\[2mm]
J_{12}&=& -4\int _0^{2\pi}d\varphi\int _{q\epsilon /h_0}^{x_{\epsilon}}\frac{x~dx}{\big (1-2xw(\vartheta,\varphi )+\sqrt{(1-2xw)^2-4x^2}\big )\sqrt{(1-2xw)^2-4x^2}}~.\nonumber
\eea
The presence of the function $w(\vartheta (h),\varphi )\vert_{h=\frac{q\epsilon}{x}}$ needs some special
care. We split the $x$-integration interval into 
\beq
I_{\mbox{\tiny lower}}~=~\big (q\epsilon /h_0,(q\epsilon )^{\delta}\big )~,~~~~I_{\mbox{\tiny upper}}~=~\big ((q\epsilon )^{\delta}, x_{\epsilon}\big )~,~~~ \mbox{with}~~~~\frac{2}{3}<\delta<1 ~.\label{intervals}
\eeq
$w(\vartheta (h),\varphi )\vert_{h=\frac{q\epsilon}{x}}$ is bounded. Furthermore, in the lower interval $I_{\mbox{\tiny lower}}$ the variable $x$ is small and we can use
\beq
\log\Big (1-2xw+\sqrt{(1-2xw)^2-4x^2}~\Big )~=~\log2~-~2xw~+~{\cal O}(x^2)~.\label{logw-what}
\eeq
This leads to \footnote{The ${\cal O}(x^2) $ term in \eqref{logw-what}, together with
the explicit prefactor $1/\epsilon ^2$, yields a contribution vanishing for $\epsilon\rightarrow 0$. This is guaranteed by the choice $\delta >2/3$ in \eqref{intervals}. }
\bea
J_{11}&=&2\pi~\log 2~\Big ((q\epsilon)^{2\delta -2}~-~\frac{1}{h_0^2}\Big )~+~o(1)\label{J11mod}\\
&&+~\frac{2}{q^2\epsilon ^2}~\int _0^{2\pi}d\varphi\int _{(q\epsilon)^{\delta}}^{x_{\epsilon}} x~\log \Big (1-2xw(\vartheta,\varphi)+\sqrt{(1-2xw)^2-4x^2}~\Big ) dx~.\nonumber
\eea
In the remaining $x$-interval $I_{\mbox{\tiny upper}}$ the angle $\vartheta$ is close to $\Omega$, since
$h=q\epsilon /x$ is small. With \eqref{sintheta}, \eqref{w} and \eqref{what} we get
\bea
w(\vartheta,\varphi )&=&\hat w(\varphi )~+~K(\varphi )~ h^2~+~\dots \nonumber\\
K(\varphi )&=&\frac{1}{4}~\big (\cos\alpha ~\cos \Omega ~+~\sin\alpha ~\cos\Omega ~\cot\Omega \big ) ~.\label{K}
\eea  
Then the expansion of the logarithm in \eqref{J11mod} yields
\bea
\log \Big (1-2xw(\vartheta,\varphi)+\sqrt{(1-2xw)^2-4x^2}~\Big )~~~~~~~~~~~~~~~~~~~~~~~~~~~~~~~~~~~~~~~~~~~~~~~~~~\\[2mm]
=~\log \Big (1-2x\hat w(\varphi)+\sqrt{(1-2x\hat w)^2-4x^2}~\Big )~-~\frac{2x~K(\varphi)~h^2}{\sqrt{(1-2x\hat w )^2-4x^2}}~+~\dots ~.\nonumber 
\eea
This estimate will be used for the further evaluation of \eqref{J11mod}, together with \eqref{xeps} and
\beq
 \int _{(q\epsilon)^{\delta}}^{x_{\epsilon}} dx~=~\left ( \int _{0}^{1/B_1}- \int _0^{(q\epsilon)^{\delta}}-\int _{x_{\epsilon}}^{1/B_1}\right )dx ~.
\eeq
Using in addition the explicit integral
\bea
\int _0^{\frac{1}{2(1+\hat w)}} x~\log \Big (1-2x\hat w +\sqrt{(1-2x\hat w)^2-4x^2}~\Big )~dx~~~~~~~~~~~~~~~~~~~~~~~~~~~~~~~~~~\\
=~\frac{1}{8(1-\hat w^2)}-\frac{1}{16(1+\hat w)^2}
-\frac{\hat w~\mbox{arccos}~\hat w}{8(1-\hat w^2)^{3/2}}-\frac{\log (1+\hat w)}{8(1+\hat w)^2}~,\nonumber
\eea
we get finally
\bea
J_{11}&=&\frac{2}{q^2\epsilon ^2}~\int _0^{2\pi}d\varphi ~\Big (
\frac{1}{8(1-\hat w^2)}-\frac{1}{16(1+\hat w)^2}
-\frac{\hat w~\mbox{arccos}~\hat w}{8(1-\hat w^2)^{3/2}}-\frac{\log (1+\hat w)}{8(1+\hat w)^2} \Big ) \nonumber \\[2mm]
&&+ 2\int _0^{2\pi}\frac{B_3(\varphi)}{B_1^3(\varphi )}~\log(1+\hat w(\varphi ))~d\varphi
\nonumber\\[2mm]
&&-~4\int _0^{2\pi}d\varphi\int _0^{1/B_1}\frac{K(\varphi )}{\sqrt{(1-2x\hat w)^2-4x^2}}~dx ~
-~\frac{2\pi~\log 2}{h^2_0}~+~o(1)~.\label{J11final}
\eea

In the asymptotic evaluation of $J_{12}$ one performs steps analogous to the above procedure. 
But, since in contrast
to  $J_{11}$, there is no divergent $1/\epsilon ^2$-prefactor, the replacement of the integration
interval for $x$ by $(0,1/B_1)$  and of $w(\vartheta ,\varphi )$ by $\hat w(\varphi )$ generates
an $o(1)$-type error only. 

Therefore, with the integral ( for $B_1~=~2(1+\hat w)$ )
\beq
\int _0^{1/B_1}\frac{dx}{\Big (1-2x\hat w+\sqrt{(1-2x\hat w)^2-4x^2}\Big )\sqrt{(1-2x\hat w)^2-4x^2}}
~=~\frac{1}{2}~,
\eeq
we get
\beq
J_{12}~=~-4\pi ~+~o(1)~.\label{J12final}
\eeq
\vspace{2mm}

Let us now turn to $J_2$, defined in \eqref{J1J2}. We again use the substitution $x=q\epsilon/h$ and
expand the integrand in $\epsilon$. To handle the divergence of the integrand at $x=0$, 
we subtract and add $1/x\cdot\log2$. This means
\bea
J_2&=& -\frac{1}{4}\int _0^{2\pi}d\varphi\int_{q\epsilon/h_0}^{x_{\epsilon}}\frac{d x}{x}~\log\Big (\frac{1-2xw(\vartheta,\varphi)+\sqrt{(1-2xw)^2-4x^2}}{2}~\Big )\nonumber\\[2mm]
&&-\frac{1}{4}\int _0^{2\pi}d\varphi\int_{q\epsilon/h_0}^{x_{\epsilon}}\frac{d x~\log 2}{x}~+~o(1)~.
\eea
Now the first term can be handled as the integral for $J_{12}$ above. In the second term
the $x$-integration is trivial, and expanding its result in $\epsilon$ we get with \eqref{xeps}
\bea
J_2&=& -\frac{1}{4}\int _0^{2\pi}d\varphi\int_{0}^{1/B_1}\frac{d x}{x}~\log\Big (\frac{1-2x\hat w(\varphi)+\sqrt{(1-2x\hat w)^2-4x^2}}{2}~\Big )\nonumber\\[2mm]
&&+\frac{\pi ~\log 2}{2}~\log (q\epsilon)~-~\frac{\pi ~\log 2}{2}~\log h_0\nonumber\\[2mm]
&& +~\frac{\log 2}{4}\int _0^{2\pi}\log B_1(\varphi )d\varphi ~+~o(1)~.\label{J2final}
\eea
\section*{Appendix B}
In ref.\cite{Drukker:2000rr} there has been
observed an anomaly under certain conformal transformations for $d=2$, see also \cite{Dorn:2013ita}. As mentioned in the introduction, for smooth $\partial A$ in $d=2$ the finite piece of $V(\gamma _A)$ is conformally invariant. This invariance is broken if $\partial A$ is a compact
contour passing the origin and is mapped under inversion on the unit circle to a contour extending 
up to infinity. The prototype is the map of circles to straight lines.

We can observe an analogous anomaly in $d=3$. There the coefficient
of the logarithmic divergence is invariant under non-exceptional conformal
transformations.\footnote{We call a conformal transformation of a compact $\partial A$ exceptional if its image is noncompact.} Let us consider spheres $S^2\subset {\mathbb R^3}$, touching the origin, which are mapped to two-dimensional planes under inversion on the unit
sphere. For the plane one gets, with an IR cut-off $l$,
\beq
V_{\epsilon}^{\mbox{\tiny plane}}~=~\frac{l^2}{2}~\frac{1}{\epsilon ^2}~.\label{plane}
\eeq
On the other side, if $\partial A$ is a sphere of radius $L$, we get 
from \cite{Ryu:2006ef} 
\beq
V_{\epsilon}^{\mbox{\tiny sphere}}~=~2\pi L^2~\frac{1}{\epsilon ^2}~+~2\pi~\log\frac
{\epsilon}{L}~+~{\cal O}(1)~.\label{sphere}
\eeq
Therefore we find an anomaly of $a$, the coefficient of the $\log$-term in \eqref{odd-GW}. It changes from zero for the planar case to $2\pi$ in the spherical case. Remarkably, it is just the same value as for the anomaly of the finite piece in the case $d=2$. 


\end{document}